# A Method for Evaluation of Aerodynamic Lift and Drag Based on Statistical Mechanics

Haibing Peng*

NOR-MEM Microelectronics Co., Ltd., Suzhou Industrial Park, Suzhou, JiangSu Province 215000, P.R. China

*Permanent Email: haibingpeng@post.harvard.edu

ABSTRACT

Despite intensive applications of Navier-Stokes equations in computational-fluid-dynamics (CFD) to understand aerodynamics, fundamental questions remain open since the statistical nature of discrete air molecules with random thermal motion is not considered in CFD. Here we introduce an approach based on Statistical Mechanics, termed as "Volume-Element" method, for numerical evaluation of aerodynamic lift and drag. Pressure and friction as a function of angle of attack have been obtained for canonical flat-plate airfoils, and the method is applicable to convex-shape airfoils directly and viable for concave-shape airfoils if combined with Monte Carlo simulations. This approach opens a door not only for aerodynamic applications, but also for further applications in Boson or Fermi gases.

## I. Introduction

Quantitative understanding of the aerodynamic lift and drag of moving objects has been under intensive research focus ever since the modern aviation age started [1-5], and continues to attract new attentions with emerging research progresses in flying biological species and micro-aerial vehicles.[4, 6-9] The mainstream practical tools of aerodynamics have been wind-tunnel experiments and computational-fluid-dynamics (CFD) simulations. With air being treated as a fluid continuum and Newton's second law being applied, Navier-Stokes (NS) equations coupled with the continuity equation have been obtained; furthermore, due to tremendous mathematical difficulty in solving the NS equations, various approximations and assumptions of boundary conditions have been applied in CFD to develop numerically solvable aerodynamics models for understanding the aerodynamic lift and drag, in which viscosity and vortex are believed to play important roles.[1-2] In history, Newton's sine-squared law [2, 10] has been obtained by treating air as a medium consisting of particles with uniform velocity striking a flying object. However, it is well known from Statistical Mechanics that air consists of an assembly of discrete individual molecules with random thermal motion at an average speed about the speed of sound $V_s$ (e.g., $V_s$ ~340 $m/s$ at room temperature), which raises the question of whether the CFD approach of treating air as a fluid continuum with *macroscopic motion* is fundamentally accurate enough no matter how small a spatial grid cell is used in CFD numerical simulations. For example, an obvious paradox is that CFD could not account for the static pressure on a closed container at rest in the laboratory frame with air inside, where the enclosed air does not have any macroscopic center-of-mass motion at all.

In this work, we introduce an approach based on Statistical Mechanics and elastic scattering of discrete molecules at the air-solid interface to address the aerodynamic lift and drag by evaluating the pressure and friction on moving objects in air. With a unique technique termed herein as "Volume-Element" method, analytical expressions of the normal-force pressure as a function of angle of attack, moving speed and temperature have been obtained for the canonical flat-plate airfoil, applicable for both ideal gases and air with intermolecular scattering; d'Alembert paradox is fully resolved and buoyant force can also be accounted for with gravitational energy considered in the statistical distribution. Furthermore, friction is found to be a direct consequence of the surface roughness of the airfoil and the relation to viscosity theory in terms of the origin of the "no-slip" boundary condition and the boundary layer is analyzed. The "Volume-Element" method developed here is applicable to any convex-shape airfoils directly and should be also viable for concave-shape airfoils if combined with Monte Carlo simulations.[11-12]

## II. Pressure on moving airfoils

### II.1. Pressure on an ideally flat plate with infinitely large surface area

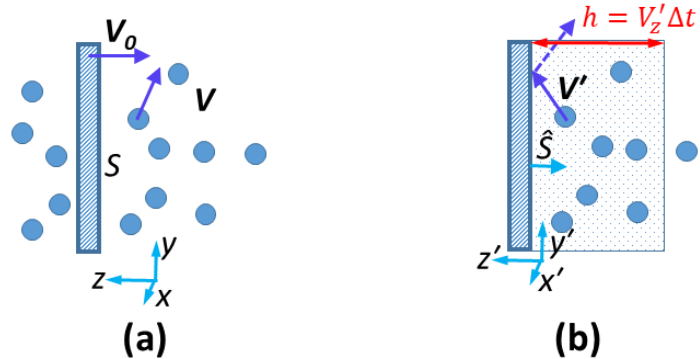

**Fig. 1.** Schematic diagrams of an infinitely large flat plate $S$ moving with velocity $V_0$: **(a)** in the laboratory frame, and **(b)** in a reference frame fixed at the plate.

We start by considering an ideal flat-plate with infinitely large surface area $S \to \infty$ and moving with a constant velocity $V_0$ perpendicular to the plate while air is at rest (*i.e.*, with no center-of-mass motion) in the laboratory frame (Fig. 1a). The velocity $V$ (in the laboratory frame) of individual air molecules is characterized by Boltzmann distribution [13] for ideal gas at temperature $T$: the probability for an molecule to have a velocity $V$ within an infinitesimal velocity-space volume $d^3V$ is $\rho(V)d^3V$, where the phase space probability density

$$\rho(V) = e^{-\beta m V^2/2}/Z, \qquad (1)$$

with $m$ being the mass of molecules, $V$ the magnitude of velocity, $\beta = 1/k_B T$ with $k_B$ the Boltzmann constant, and the partition function $Z = (2\pi/\beta m)^{3/2}$.

On the front side of the plate, the surface normal $\hat{S}$ is parallel to the plate velocity $V_0$. For convenience, we choose a reference frame fixed at the moving plate with $z'$ axis antiparallel to $V_0$ (Fig. 1b), wherein the molecule velocity is now $V' = V - V_0$, and according to eq. (*1*) the related probability density is now

$$\rho_1(V') = \rho(V) = \rho(V' + V_0) = e^{-\beta m |V' + V_0|^2 / 2}/Z. \tag{2}$$

To characterize the scattering of molecules at the air-plate interface, we first take the case of a plate with *ideally flat* surfaces, e.g., atomically flat crystal surfaces or molecular-level flat surfaces formed by adsorption of molecules; and the plate mass $M$ is much larger that the molecular mass $m$. We assume that the scattering is elastic, *i.e.*, assume that the probability of non-elastic scattering events (*e.g.*, phonon excitation or absorption in solid) is small and such non-elastic scattering induced effect as a statistical average can be neglected. We further treat the elastic scattering of air molecules by the plate as ping-pong balls bouncing back from a rigid wall, *i.e.*, the molecular velocity component along the surface tangential is unchanged after the scattering while the velocity component along the surface normal is changed accordingly based on energy and momentum conservation; this is reasonable since for an ideally flat plate the repulsive force perpendicular to the surface should be far more stronger than the force parallel the surface, so that the impact and thus the change of momentum are dominant in the direction perpendicular to the surface.

In the reference frame fixed at the plate, a single Ping-Pang-ball-like elastic scattering event (Fig. 1b) with an incident molecule velocity $V'$ causes a momentum change of the individual molecule along $z'$ axis and its magnitude is given by: $\Delta P = 2mV_z'$, since the plate mass $M \gg m$.

At any given time *t*, all molecules with any given $V_z' > 0$ (*i.e.*, moving towards the plate) in a space region within a distance $h = V_z' \Delta t$ away from the plate along the surface normal $\hat{S}$, as shown in Fig. 1b, are able to reach the infinitely large plate within a time interval $\Delta t$, and each molecule induces an average normal force on the plate as: $f = \Delta P/\Delta t = 2mV_z'/\Delta t$ according to Newton's second and third laws. Given a molecule density $n$ in air, the total number of molecules in the shaded space region (Fig. 1b) within a distance $h = V_z' \Delta t$ is $N = n \cdot Sh = nSV_z' \Delta t$. Also, considering the statistical homogeneity of the spatial distribution of air molecules and the symmetry of their velocity distribution, we know that the pressure (*i.e.*, normal force per unit surface area) acting on the flat plate ($S \rightarrow \infty$) is uniform all over the surface. Summing up such normal forces induced by all molecules within the shaded space regions associated with all values of $V_z' > 0$ and using $V_z' = V_z + V_0$, we have the pressure acting on the front side of the plate as:

$$p_+ = \frac{\int_{-\infty}^{+\infty} dV_x' \int_{-\infty}^{+\infty} dV_y' \int_0^\infty \rho_1(V') \cdot (2mV_z'/\Delta t) \cdot nSV_z' \Delta t \ dV_z'}{S}$$

$$= \int_{-\infty}^{+\infty} dV_x \int_{-\infty}^{+\infty} dV_y \int_{-V_0}^{\infty} \rho(V) \cdot 2nm(V_z + V_0)^2 \ dV_z, \tag{3}$$

wherein we have changed the integration variables back to $V$ space and $\rho(V)$ is given by eq. (*1*). By integrating out the variables $V_x$ and $V_y$ in eq. (*3*), we further have:

$$p_+ = \frac{2nm}{\sqrt{2\pi/\beta m}} \int_{-V_0}^{\infty} e^{-\beta m V_z^2/2} \cdot (V_z + V_0)^2 \, dV_z$$

$$= \frac{2nm}{\sqrt{2\pi/\beta m}} \{\int_{-V_0}^{0} e^{-\frac{\beta m V_z^2}{2}} \cdot (V_z^2 + 2V_z V_0 + V_0^2) \, dV_z + \int_{0}^{\infty} e^{-\frac{\beta m V_z^2}{2}} \cdot (V_z^2 + 2V_z V_0 + V_0^2) \, dV_z\}.$$

Finally, using integral-evaluation techniques with parametric differentiation under the integral sign, we then obtain an analytic expression of the pressure on the front side of the plate as:

$$p_+ = \left(\frac{n}{\beta} + nmV_0^2\right) \cdot \left\{1 + \mathrm{erf}\left(\sqrt{\frac{\beta m}{2}} V_0\right)\right\} + \sqrt{\frac{2}{\pi \beta m}} \cdot nmV_0 \cdot e^{-\frac{\beta m V_0^2}{2}}$$

$$= (p_0 + nmV_0^2) \cdot \left\{1 + \mathrm{erf}\left(\sqrt{\frac{\beta m}{2}} V_0\right)\right\} + \frac{2}{\sqrt{\pi}} p_0 \cdot \sqrt{\frac{\beta m}{2}} V_0 \cdot e^{-\frac{\beta m V_0^2}{2}}. \quad (4)$$

Here $p_0 = \frac{n}{\beta} = nk_B T$ is the well-known static pressure for ideal gas at rest inside a container, and the error function is defined as $\mathrm{erf}(x) \equiv \frac{2}{\sqrt{\pi}} \int_0^x e^{-t^2} dt$.

Following similar procedures as described above, we can also obtain the pressure on the back side of the plate as:

$$p_- = (p_0 + nmV_0^2) \cdot \{1 - \mathrm{erf}(\sqrt{\frac{\beta m}{2}} V_0)\} - \frac{2}{\sqrt{\pi}} p_0 \cdot \sqrt{\frac{\beta m}{2}} V_0 \cdot e^{-\frac{\beta m V_0^2}{2}}. \quad (5)$$

**II.2. Pressure on an ideally flat plate with finite surface area**

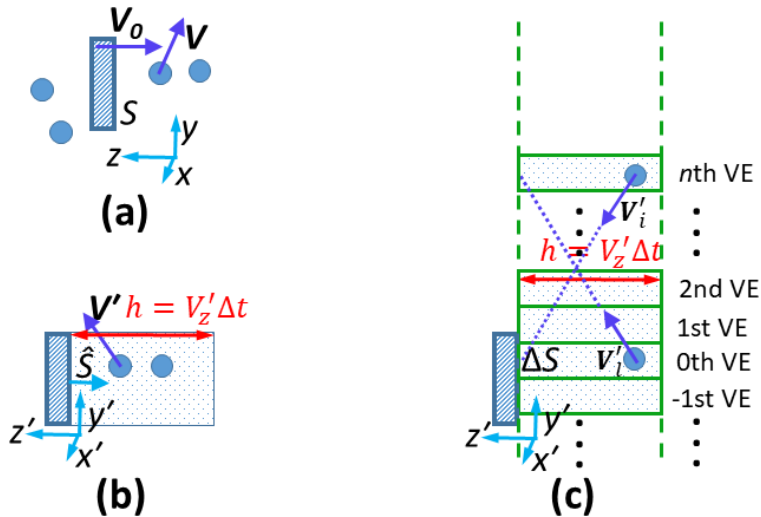

**Fig. 2.** Schematic diagrams of a finite flat plate $S$ moving with velocity $V_0$ perpendicular to the plate: **(a)** in the laboratory frame, **(b)** in a reference frame fixed at the plate, and **(c)** with a set of VEs constructed for a given $V_z'$.

Now we consider an ideal flat-plate airfoil with finite surface area $S$, moving with a constant velocity $\mathbf{V_0}$ perpendicular to the plate in the laboratory frame (Fig. 2a). Working in the reference frame fixed at the moving plate (Fig. 2b), for molecules with a given $V_z' > 0$ we construct a shaded space region within a distance $h = V_z'\Delta t$ away from the plate with finite $S$.

However, different than the situation for an infinitely large plate of Fig. 1b, the previous reasoning applied to obtain eq. (*3*) is no longer justified for the plate of Fig. 2b with finite area, since at any given time $t$, some molecules with a given $V_z' > 0$ in the shaded space region of Fig. 2b are not able to reach the plate with finite $S$ if the molecular velocity direction is not pointing towards the plate area.

Nevertheless, we show that eq. (*3*) is still applicable for any flat-plate with finite surface area, using the Volume-Element method described below.

First, we take a small surface element $\Delta S = \Delta x' \Delta y'$ in the plate, and for molecules with a given $V_z' > 0$ we construct the zeroth volume element (VE) as the space region within a distance $h = V_z'\Delta t$ away from $\Delta S$. Next, by periodic translation of the surface element $\Delta S$ and its associated VE along the infinitely large surface plane containing the plate and beyond, a set of VEs associated with the given $V_z'$ are formed (Fig. 2c) and every VE is filled with air molecules with $V_z'$ fixed but $V_x'$ and $V_y'$ unrestricted. With $V_z'$ values taken from 0 to ∞ and one corresponding set of VEs for every $V_z'$ constructed, all air molecules moving towards the plate plane and being able to reach the plane within a time interval $\Delta t$ are included in these complete sets of VEs.

Note that the surface element $\Delta S$ can even be macroscopically small enough to be treated as infinitesimal as long as each VE is statistically homogeneous and the velocity of molecules therein satisfies Boltzmann distribution. We further assume that the effective air temperature $T$ characterizing every VE is the same so that the pressure is uniform all over the plate surface $S$ (the effect of plate motion on air temperature and non-uniform pressure distribution will be discussed later).

Considering the symmetry of the probability distribution shown in eq. (*2*) with $\mathbf{V_0}$ antiparallel to $z'$ axis, one can infer that for any molecule with velocity $\mathbf{V_l'}$ outgoing from the zeroth VE to the *n*th VE there must exist a *statistically* pairing molecule with velocity $\mathbf{V_i'}$ incoming from the *n*th VE to the zeroth VE (Fig. 2c) where $\mathbf{V_l'}$ and $\mathbf{V_i'}$ have the same $z'$ component $V_z'$ but their $x'$ ($y'$) components have the same magnitude but opposite directions; and *vice versa* for any incoming molecule with velocity $\mathbf{V_i'}$, there exists a *statistically* pairing outgoing molecule with velocity $\mathbf{V_l'}$.

Upon reaching the plate surface of the relevant VE, each pairing molecule contributes the same amount of momentum transfer of $\Delta P = 2mV_z'$ and thus induces the same average normal force on the plate. Such a one-to-one correspondence between the incoming and outgoing molecules leads to an interesting result: the total momentum transfer on the surface element $\Delta S$, induced by those molecules physically incident from all VEs, is equal to that induced by all molecules contained in the zeroth VE (no matter whether the molecules are actually travelling out of the zeroth VE or not).

Therefore, at any given time $t$, it can be treated *effectively* as if all molecules in the zeroth VE associated with a given $V_z' > 0$, defined as the space region within a distance $h = V_z'\Delta t$ away from $\Delta S$, were able to reach the infinitesimal surface element $\Delta S$ within a time interval $\Delta t$, and each molecule induces an effective average normal force on $\Delta S$ as: $f = \Delta P/\Delta t = 2mV_z'/\Delta t$.

Summing up such normal forces "induced" by all molecules within the zeroth VEs associated with all values of $V_z' > 0$ via integration, we again reach eq. (*3*) [14] and consequently obtain the pressure on the front and back side of the surface element $\Delta S$ as those given by eqs. (*4*) and (*5*), respectively.

With eqs. (*4*) and (*5*), the net pressure (i.e., the net normal force per unit area) acting on the finite plate as a result of the plate motion can be obtained as

$$p_{net} \equiv p_+ - p_- = 2(p_0 + nmV_0^2) \cdot \text{erf}\left(\sqrt{\frac{\beta m}{2}} V_0\right) + \frac{4}{\sqrt{\pi}} p_0 \cdot \sqrt{\frac{\beta m}{2}} V_0 \cdot e^{-\frac{\beta m V_0^2}{2}}. \qquad (6)$$

In general, one can use tabulated values of the error function erf (*x*) for exact evaluation of eqs. (*4*)-(*6*), or use converging Bürmann series [15-16]:

$$\text{erf}(x) = \frac{2}{\sqrt{\pi}} \text{sgn}(x) \cdot \sqrt{1 - e^{-x^2}} \left(\frac{\sqrt{\pi}}{2} + \sum_{j=1}^{\infty} C_j e^{-jx^2}\right),$$

where sgn(*x*) is the sign function, and as a good approximation the first two expansion terms can be used as $C_1 = 31/200$ and $C_2 = -341/8000$.

Below we discuss the results for a few limiting cases. (**1**) If the plate is at rest, $V_0 = 0$, we have $p_{net} = 0$, and $p_+ = p_- = p_0$, the pressure for ideal gas. (**2**) For finite $V_0$ at temperature $T = 0$, $p_{net} = 2nmV_0^2$, which can be explained straightforwardly by Newton's second law since without thermal motion each molecule contributes a momentum transfer of $2mV_0$ to the moving plate and the number of molecules colliding with the plate per unit area within a time interval $\Delta t$ is equal to $nV_0\Delta t$. (**3**) For finite $V_0$ at low temperature limit where $\frac{mV_0^2}{2k_BT} \gg 1$ (this is equivalent to the high-Mach-number case $V_0 \gg V_s$, with the speed of sound $V_s = \sqrt{k_BT/m}$ for idea gas ), we have $p_{net} \approx 2(p_0 + nmV_0^2)$. (**4**) For finite $V_0$ at high temperature limit where $\frac{mV_0^2}{2k_BT} \ll 1$ (this is equivalent to the low-Mach-number case $V_0 \ll V_s$), we have

$$p_{net} \approx \frac{4}{\sqrt{\pi}} (2p_0 + nmV_0^2) \cdot \sqrt{\frac{\beta m V_0^2}{2}} \cdot e^{-\frac{\beta m V_0^2}{2}},$$

where the first-order Taylor expansion of the error function erf (*x*) has been used.

Furthermore, we discuss the effect of plate motion on the effective air temperature profile *in close proximity to* the plate surface. If the plate is at rest, in the laboratory frame, the statistical distribution of air molecules in velocity space is spherically symmetric, filled up to a cutoff speed $V_c$ (Fig. 3a); this is the same as the case without the presence of the plate since elastic scattering simply reverses the semi-sphere of incident velocity distribution with $V_z > 0$ into the semi-sphere of outgoing velocity distribution with $V_z < 0$.

If the plate starts moving at **V₀**, in the laboratory frame, the velocity distribution of molecules at the plate surface is disturbed instantaneously due to elastic scattering between molecules and the plate: as described in Figs. 3b-3e, the portion of incident velocity distribution (unshaded area) is mirror reflected into the portion of outgoing velocity distribution (shaded area) as a consequence of elastic scattering, and the incident and the outgoing portions together form the complete instantaneous velocity distribution for molecules at the plate surface (see Figs.

3b and 3d for the cases on the front side of the plate, and Figs. 3c and 3e for the cases on the back side).

However, if the outgoing molecules cause further intermolecular collisions, the velocity distribution of surrounding space may reach a local quasistatic-equilibrium state which can be described by an effective temperature $T_e$ and an effective center-of-mass displacement velocity $\boldsymbol{V}_e = -V_e\boldsymbol{z}$ (Fig. 3f).

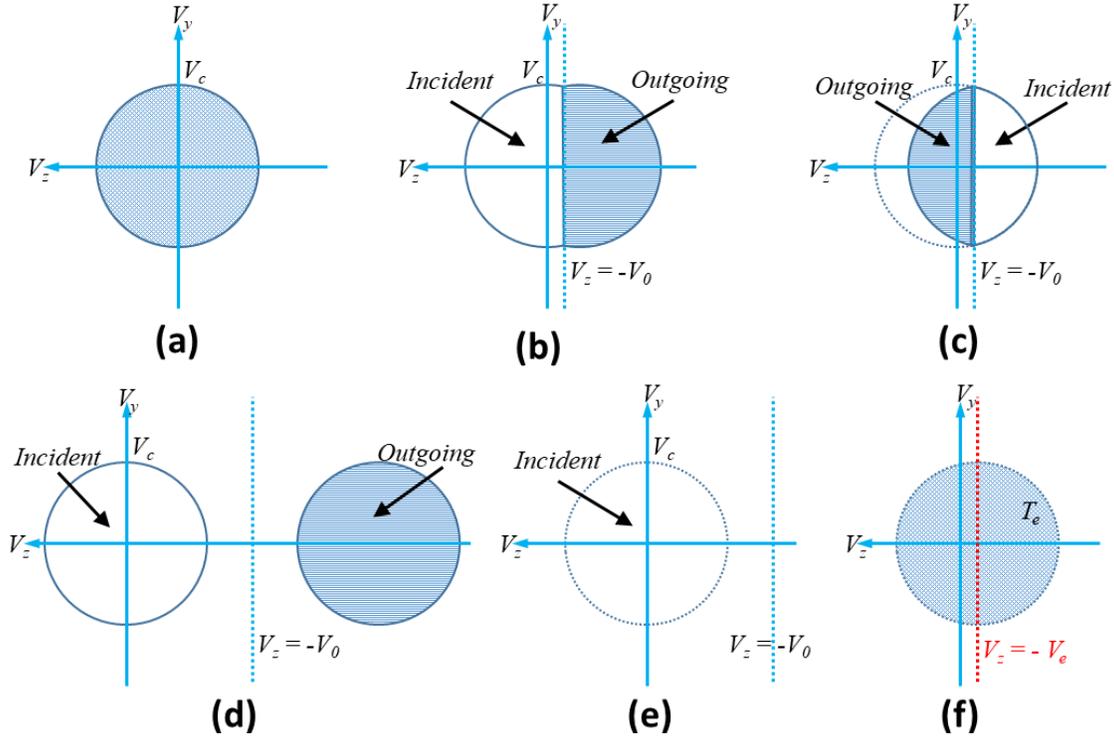

**Fig. 3.** (a) Statistical distribution of molecules in velocity space for a plate at rest. (b-e) Instantaneous statistical distribution of molecules in velocity space for a moving plate with $V_0 < V_s$: (b) on the front side and (c) on the back side of the plate surface, respectively; and with $V_0 > V_s$: (d) on the front side and (e) on the back side of the plate surface, respectively. (f) Statistical distribution of molecules in velocity space described by a quasistatic-equilibrium state with an effective temperature $T_e$ and an effective center-of-mass displacement velocity $\boldsymbol{V}_e = -V_e\boldsymbol{z}$.

Depending on the magnitude of $\boldsymbol{V_0}$, the plate length $l_{plate}$ and the mean free path $l_{mfp}$ characterizing the intermolecular collisions, the steady state of the velocity distribution for the surrounding space should be somewhere between the instantaneous cases of Figs. 3b-3e and the quasistatic-equilibrium state of Fig. 3f, and such a steady state distribution is the velocity distribution actually sensed by the plate when it keeps moving at $\boldsymbol{V_0}$. Therefore, the pressure on the plate can be obtained by directly applying eqs. (4)-(6) if the velocity distribution actually sensed by the plate is described by Figs. 3b-3e where the effect of intermolecular collisions is negligible, or by modifying eqs. (4)-(6) via replacing $T$ with $T_e$ and $V_0$ with $V_0' = V_0 - V_e$ therein if the velocity distribution actually sensed by the plate is described by Fig. 3f where a local quasistatic-equilibrium state is reached.

Estimation of $T_e$ and $V_e$ are given below for a few limiting cases:

**(1)** For $V_0 \ll V_s$ and $l_{plate} \ll l_{mfp}$, the effect of intermolecular collisions is negligible, so that the velocity distribution sensed by the plate is the semi-sphere of incident velocity distribution in Figs. 3b and 3c, and thus $T_e$ is equal to the undisturbed air temperature $T$ and $V_e = 0$.

**(2)** For $V_0 < V_s$ and $l_{plate} \gg l_{mfp}$, intermolecular collisions causes further thermalization, so that the velocity distribution sensed by the plate should be the quasistatic-equilibrium states of Fig. 3f, with $V_e \sim V_0$ but $T_e > T$ on the frontside and $T_e < T$ on the backside of the plate, respectively, as can be inferred from the center-of-mass speed and average molecular energy since kinetic energy and momentum of the system are conserved in intermolecular collisions which causes the evolution of the states of Fig. 3b or 3c into a thermalized quasistatic-equilibrium state described by Fig. 3f.

**(3)** For $V_0 \gg V_s$, on the front side of the plate, thermalization likely occurs rapidly as a result of multiple intermolecular collisions of any individual supersonic molecule until its energy is significantly reduced, and the actual velocity distribution sensed by the plate on the front side may be in an equilibrium state of Fig. 3f with $V_e \sim 0$ but $T_e \sim T+\Delta T$, where $\Delta T \sim \frac{4}{3} m V_0^2 / k_B$ as estimated by converting the gain of kinetic energy $\sim \frac{1}{2} m (2V_0)^2$ per molecule from scattering at the plate into an increasement of average thermal energy $\frac{3}{2} k_B \Delta T$ due to thermalization of high-speed molecules ($\sim 2V_0$) via collisions with low-speed molecules from environmental air at temperature $T$.

**(4)** For $V_0 \gg V_s$, on the back side of the plate, the plate experiences few scattering from air molecules (Fig. 3e) and the resultant pressure $p_-$ is small but can still be described by eq. (*5*) since certain kind of instantaneous vacuum state may be formed and the effective velocity distribution can be determined by the diffusion of molecules from these two ends of the plate.[17]

In general cases, $T_e$ and $V_e$ can have a spatial dependence, and the total momentum transfer on the surface element $\Delta S$ can be numerically evaluated via Monte Carlo simulations.

For example, if $T_e$ and $V_e$ are different in different VEs or even within the same VE, we can first divide each VE into a manageable number of grid cells in position space, and for each target grid cell within a VE associated with a given $V_z' > 0$ we generate molecules statistically via Monte Carlo method with random velocity components $V_x'$ and $V_y'$ (with their magnitude up to a reasonable cut-off speed $V_c$) using the probability distribution with $T_e$ and $V_e$ values for the target grid cell. After that, for each generated molecule, based on its velocity we can determine the specific VE where it will reach the plate plane within the time interval $\Delta t$ and also its momentum transfer to the surface element of that specific VE.

To limit the computation load, the simulations can be run over a reasonable number of VEs (associated with a given $V_z'$) within a distance away from the zeroth VE in the length scale $\sim l_{mfp}$ (the mean free path characterizing intermolecular collisions) in the $x'y'$ plane, and also run over limited values of $V_z'$ so that the distance $h = V_z' \Delta t$ is in the length scale $\sim l_{mfp}$ to limit the set of VEs need to be considered.

Finally, summing up the total momentum transfer on the surface element $\Delta S$ by the molecules incident from all VEs simulated, we then obtain the normal force and thus the pressure on the surface element numerically. The above numerical method via Monte Carlo simulations should be of practical use for analysis of complicated situations if combined with experimental measurements to determine the spatial dependence of $T_e$ and $V_e$ to be used for simulations.

Additionally, one could also theoretically estimate $T_e$ and $V_e$ by treating the rebounding of molecules at the plate surface as a perturbation and using the linear response theory recently developed.[17]

### II.3. Pressure-induced lift and drag on an ideally flat plate with angle of attack

Next, we consider an ideal flat-plate airfoil with an area $S$ and an angle of attack (Fig. 4a), moving in the laboratory frame with a constant velocity $\boldsymbol{V_0}$ at an angle $\theta_0$ from the plate surface normal $\hat{\boldsymbol{S}}$ (in literatures, the angle of attack is usually defined as $\alpha = \frac{\pi}{2} - \theta_0$). Similar to the method described in Sec. **II.2**, working in the reference frame fixed at the moving plate, we take a small surface element $\Delta S = \Delta x' \Delta y'$ in the plate, and for molecules with a given $V_z' > 0$ we construct a set of VEs within a distance $h = V_z' \Delta t$ away from the plate (Fig. 4b). Again, we assume that the effective air temperature $T$ characterizing every VE is the same so that the pressure is uniform over the plate surface $S$ (the effect of plate motion on air temperature and non-uniform pressure distribution is similar to the discussion of Sec. **II.2**).

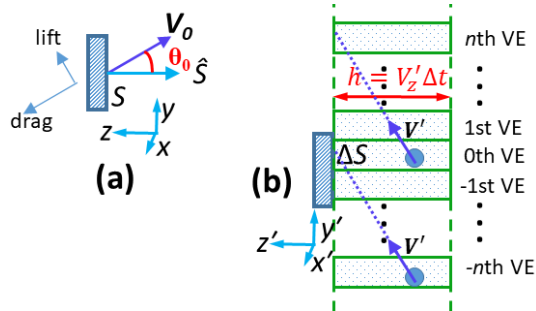

**Fig. 4.** Schematic diagrams of a finite flat plate $S$ moving with velocity $\boldsymbol{V_0}$ at an angle $\theta_0$ from the plate surface normal $\hat{\boldsymbol{S}}$: **(a)** in the laboratory frame, and **(b)** in a reference frame fixed at the plate with a set of VEs constructed for a given $V_z'$.

Since the probability distribution of eq. (*2*) depends only on the velocity $\boldsymbol{V'}$ but is translationally invariant in position space, for a molecule with any given velocity $\boldsymbol{V'}$ outgoing from the zeroth VE to the $n$th VE there must exist a *statistically* pairing molecule with the same velocity $\boldsymbol{V'}$ incoming from the -$n$th VE to the zeroth VE (Fig. 4b), considering that the $n$th VE and -$n$th VE are related to the zeroth VE via translational symmetry in position space; upon reaching the plate surface, each pairing molecule contributes the same amount of momentum transfer of $\Delta P = 2mV_z'$ and thus induces the same average normal force on the plate.

Again, such a one-to-one correspondence of the incoming and outgoing molecules leads to the result: the total momentum transfer on the surface element $\Delta S$ of the zeroth VE, induced by those molecules physically incident from all VEs, is equal to that induced by all molecules contained in the zeroth VE (no matter whether the molecules are actually travelling out of the zeroth VE or not).

Therefore, at any given time $t$, it can be treated *effectively* as if all molecules in the zeroth VE associated with a given $V_z' > 0$, defined as the space region within a distance $h = V_z' \Delta t$ away from $\Delta S$, were able to reach the infinitesimal surface element $\Delta S$ within a time interval $\Delta t$, and each molecule induces an effective average normal force on $\Delta S$ as: $f = \Delta P/\Delta t = 2mV_z'/\Delta t$. Summing up such normal forces "induced" by all molecules within the zeroth VE associated with all values of $V_z' > 0$ and using $V_z' = V_z + V_0 \cos \theta_0$, we can follow the steps from eq. (3) to eq. (5) and obtain the pressure on the front and back side of the plate, respectively, as

$$p_+ = (p_0 + nmV_0^2 \cos^2 \theta_0) \cdot \left\{1 + \text{erf}\left(\sqrt{\frac{\beta m}{2}} V_0 \cos \theta_0\right)\right\} + \frac{2}{\sqrt{\pi}} p_0 \cdot \sqrt{\frac{\beta m}{2}} V_0 \cos \theta_0 \cdot e^{-\frac{\beta m V_0^2 \cos^2 \theta_0}{2}} \tag{7}$$

and

$$p_- = (p_0 + nmV_0^2 \cos^2 \theta_0) \cdot \left\{1 - \text{erf}\left(\sqrt{\frac{\beta m}{2}} V_0 \cos \theta_0\right)\right\} - \frac{2}{\sqrt{\pi}} p_0 \cdot \sqrt{\frac{\beta m}{2}} V_0 \cos \theta_0 \cdot e^{-\frac{\beta m V_0^2 \cos^2 \theta_0}{2}} \tag{8}$$

Note that eqs. (7) and (8) for the ideal flat-plate moving with an angle of attack, can be reached by simply replacing $V_0$ in eqs. (4) and (5) with $V_0 \cos \theta_0$, the component of the plate velocity along the direction of the surface normal $\hat{S}$. With eqs. (7) and (8), the net pressure acting on the ideal flat-plate moving with an angle of attack is obtained as

$$p_{net} \equiv p_+ - p_-$$
$$= 2(p_0 + nmV_0^2 \cos^2 \theta_0) \cdot \text{erf}\left(\sqrt{\frac{\beta m}{2}} V_0 \cos \theta_0\right) + \frac{4}{\sqrt{\pi}} p_0 \cdot \sqrt{\frac{\beta m}{2}} V_0 \cos \theta_0 \cdot e^{-\frac{\beta m V_0^2 \cos^2 \theta_0}{2}}. \tag{9}$$

Consequently, the lift force per unit area on the plate is

$$f_L = p_{net} \cdot \sin \theta_0, \tag{10}$$

and the drag force per unit area on the plate is

$$f_D = p_{net} \cdot \cos \theta_0. \tag{11}$$

The net pressure on the ideal flat-plate as a function of angle of attack given by eq. (9) is also dependent on both the temperature $T$ and the moving speed $V_0$, particularly, showing an intriguing and physically reasonable scaling behavior as a function of $\frac{mV_0^2}{k_B T} = \frac{V_0^2}{V_S^2}$, with the speed of sound $V_S = \sqrt{k_B T/m}$ for idea gas. The limiting cases have already been discussed in Sec. **II.2**; and the results agree well with known physical situations such as the static pressure for ideal gas at rest, and Newton's method [2, 10] equivalent to the zero-temperature limit (so that statistical distribution can be neglected). We note that the important dependence of the net pressure on both $T$ and $V_0$, as given by eq. (9), is an intrinsic outcome of Statistical mechanics wherein air motion is described by thermal motion uniquely characterized by temperature and by an effective center-of-mass displacement speed $V_0$ in the frame fixed at the plate. This important dependence on both $T$ and $V_0$ is absent in traditional aerodynamic theories [1, 2] which fail in capturing the essential statistical velocity distribution of thermal motion: *e.g.*, in Fig. 3a of Ref. [2], the lift coefficient is expressed only as a function of angle of attack but independent of the temperature and the plate moving speed, which is in stark contrast to the result of eq. (9) in this work. The scaling behavior as a function of $\frac{mV_0^2}{k_BT} = \frac{V_0^2}{V_s^2}$ (*i.e.*, the square of the Mach number) in eq (9) can be tested in future experiments by plotting the measured lift/drag force as a function of angle of attack at different speed (expressed as Mach numbers) with the air temperature fixed.

      Below we discuss the relation of this work to conventional aerodynamic theories.[1, 2] The vortex-based theories,[1, 2] *e.g.*, from Kutta-Joukowski circulation theory, to Prandtl's "inviscid" vortex-force theory, and further to the viscous vortex-force model, are probably the most reasonable **attempts of approximation** to mimic the statistical velocity distribution of thermal motion. The key lies in how to describe thermal motion with reasonable approximations using conventional motion concept. First, we break air (a fluid) into many small elements in position space $d\Omega \equiv d^3\boldsymbol{r}$, which are *macroscopically* infinitely small but *microscopically* large enough to contain many discrete molecules accountable by statistical distributions; and we take the element $d\Omega$ as characterizing the smallest length scale of interest for study. In terms of continuum Newtonian mechanics, if air is at rest in the laboratory frame, every element $d\Omega$ does NOT have any net linear momentum or net angular momentum. Therefore, with air at rest, the best approximation to describe the thermal motion of an element $d\Omega$ is to treat it as a built-in pair of vortex and anti-vortex, with zero center-of-mass displacement velocity but with opposite vorticity vectors as **ω** and -**ω**, respectively; and the direction of **ω** is random distributed in three-dimensional position space. By this assumption of pairing vortex and anti-vortex, for any volume in position space which contains a number of those "smallest" elements $d\Omega$, its net linear momentum and angular momentum are both zero, as an approximation consistent with the result based on the Boltzmann distribution; in addition, potential flow does NOT exist in air at rest since it will generate local linear momentum inconsistent with the physical result of random thermal motion (if the air as a whole is moving, it is equivalent to a movement of all pairs of vortex and anti-vortex, *i.e.*, still being a rotational flow). Moreover, we note that the density of the pairs of vortex and anti-vortex should be proportional to $e^{-\frac{I\omega^2}{k_BT}}$ according to Statistical mechanics, with $I$ being the effective moment of inertia for the vortex (or a rotating object), since the energy of a vortex (or anti-vortex) is $I\omega^2/2$; nevertheless, we take the density as its average value to be more consistent with conventional fluid mechanics.

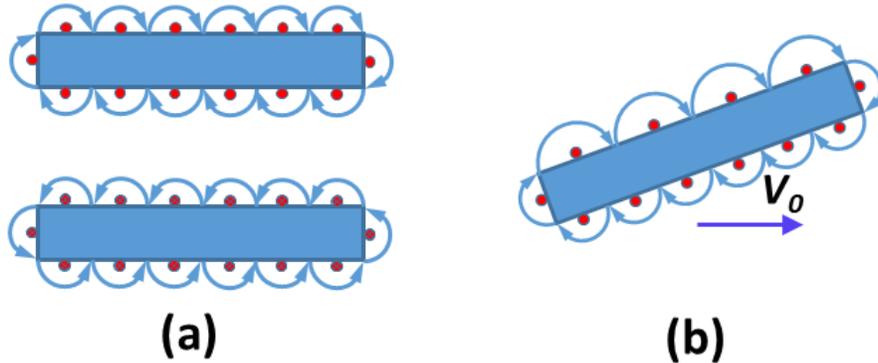

**Fig. 5.** Schematic diagrams of: **(a)** pairing edge states for semi-vortex (upper) and semi-anti-vortex (bottom) in a plate with zero angle of attack; **(b)** edge state for semi-vortex in a plate moving with finite angle of attack (the pairing counterpart edge state for semi-anti-vortex is not drawn here), wherein the density of semi-vortex in the lower and upper surface is different.

Next, we introduce an idea flat-plate into the air. The flat-plate then interacts with the pairs of vortex and anti-vortex; if only elastic scattering is considered and the velocity component perpendicular to the plate is reversed due to specular reflection, a likely steady state is that pairs of vortex and anti-vortex at air-plate interface break into semi-vortex and semi-anti-vortex so that two pairing edge states are formed around the flat-plate, one for the vortex and the other for the anti-vortex counterpart (Fig. 5). The net vortex force (or centripetal force) for each semi-vortex is now perpendicular to the plate surface (*i.e.*, a normal force), and thus the integration of vortex force along all possible edge states gives the origin of the lift and drag on the plate. If the flat-plate is at rest (Fig. 5a) or moving parallel to the plate, the density of semi-vortex or semi-anti-vortex is the same for the upper and the bottom surface of the plate, resulting in a zero net force after summing up the normal force on both surfaces. However, if the flat-plate is moving with an angle of attack (Fig. 5b), the bottom surface will interact with more pairs of vortex and anti-vortex as it is moving to catch more pairs; therefore, the effective density of semi-vortex or semi-anti-vortex on the bottom is larger than that on the upper surface, leading to a non-zero net normal force on the plate which gives lift and drag.

## II.4. Pressure-induced lift and drag on convex-shape airfoils

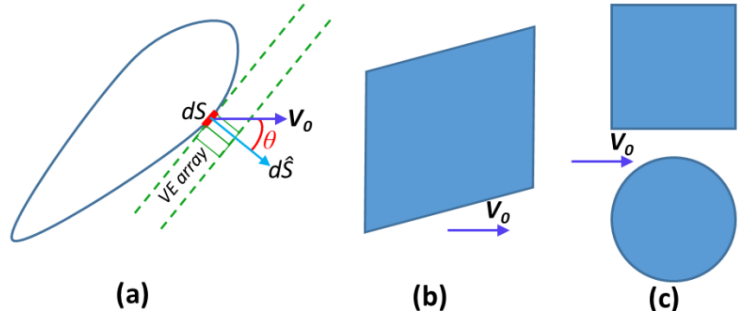

**Fig. 6.** (a) Schematic diagram of a convex-shape two-dimensional airfoil moving with velocity $V_0$, where the closed convex-shape surface consists of many infinitesimal surface elements $d\mathbf{S}$. (b) A monoclinic-crystal shape object, and (c) a cubic and a spherical solid object moving at $V_0$, respectively.

For any convex-shape airfoils, we can break the whole surface into infinitesimal surface elements $d\mathbf{S}$ and apply the Volume-Element method to obtain the local pressure and sum up the contribution of normal force acting on every surface element via integration.

Taking a two-dimensional airfoil as an example (Fig. 6a), we break the closed surface into many infinitesimal surface elements, and for each surface element $d\mathbf{S}$ we construct a set of VEs as described in Sec. **II.2** by periodic translation of $d\mathbf{S}$ and its associated VE within the infinitely large surface plane containing $d\mathbf{S}$. As described in Sec. **II.3**, the pressure acting on each surface element $d\mathbf{S}$ is expressed by eq. (7), except that the angle $\theta_0$ therein is here replaced by a continuously changing variable $\theta$ characterizing the angle between the airfoil velocity vector $\mathbf{V_0}$ and the surface normal vector $d\hat{\mathbf{S}}$ (the angle $\theta$ is taken positive if $\mathbf{V_0}$ is counter-clockwise with respect to $d\hat{\mathbf{S}}$). Since the normal force acting on each surface element is $d\mathbf{F}_N = -p_+(\mathbf{V_0},\ \theta)\, d\mathbf{S}$, we thus obtain the total net force acting on the airfoil as:

$$\mathbf{F}_{net} = \oint -p_+(\mathbf{V_0},\ \theta)\, d\mathbf{S}, \tag{12}$$

where the integration is taken over the whole closed surface of the airfoil. Note that in more general cases, e.g., when the airfoil is rotating or the air motion is complicated, the effective velocity vector $V_0$ is also dependent locally on the surface element $dS$.

Below we discuss the well-known d'Alembert paradox: *i.e.*, for incompressible ideal fluids, the net force acting on a solid object inside a potential flow is zero.[18] In the theory of this work, d'Alembert paradox is fully resolved, likely because that the built-in random thermal motion of air does not present any local potential flow at all (as previously described by the approximation with paired vortex and anti-vortex). As shown in Fig. 6b, for a monoclinic-crystal shape object moving horizontally in air with a speed $V_0$, both the lift and the drag are non-zero, which can be seen easily from eq. (9) since the monoclinic-crystal effectively consists of two equivalent flat-plates of Sec. **II.3** , with one plate perpendicular and the other at an angle to the direction of $V_0$. Also, with eq. (9) and the geometric symmetry considered, a cubic or spherical object moving horizontally gives zero lift but non-zero drag (Fig. 6c).

Moreover, the Statistical-Mechanics based theory of this work can also include the effect of gravitational potential into the Boltzmann distribution, since the phase space probability is proportional to $e^{-\beta E(V,r)} d^3V d^3r$, with $E(V,r)$ being the energy for a molecule at position $r$ and with velocity $V$. By this, for air in atmosphere near the earth surface, the unnormalized phase space probability density is:

$$e^{-\beta E(V,r)} = e^{-\beta mgH(r)} e^{-\beta mV^2/2}, \qquad (13)$$

with $H(r)$ the height above the earth surface at position $r$. Using this modified probability density, the normal force acting on the surface element of Fig. 6a is now $d\boldsymbol{F}_N = -p_+(\boldsymbol{V_0},\ \theta)\ e^{-\beta mgH(dS)} dS$ . Thus the total net force acting on the airfoil is:

$$\boldsymbol{F}_{net} = \oint -p_+(\boldsymbol{V_0},\ \theta)\ e^{-\beta mgH(dS)} d\boldsymbol{S}\ , \qquad (14)$$

where $H(d\boldsymbol{S})$ is the height above the earth surface at the surface element $d\boldsymbol{S}$. With gravity considered, one can easily show that for the case of Fig. 6c with a cube or a sphere moving horizontally, the lift force is simply Archimedes' buoyant force resultant from the mass of air dispelled by the solid object.

**II.5. The effect of intermolecular scattering on pressure**

The previous derivation of the net pressure of eq. (9) is purely based on kinetics theory without considering intermolecular scattering; thus seemingly the result should work only for air with large mean free path. However, the sufficient condition for deriving eq. (7)-(9) is: "**at any given time *t*, it can be treated *effectively* as if all molecules in the zeroth VE associated with a given $V'_z > 0$, defined as the space region within a distance $h = V'_z \Delta t$ away from $\Delta S$, were able to reach the infinitesimal surface element $\Delta S$ within a time interval $\Delta t$ .**" It is surprising but intriguing that this sufficient condition also holds for situations with intermolecular scattering as long as the length scale $l_s$ associated with the *macroscopically* "infinitesimal" surface element $\Delta S$ (which we can choose depending on the study of interest as long as $\Delta S$ is *microscopically* large enough to contain many discrete molecules accountable by statistical distributions), i.e., $l_s \equiv \sqrt{\Delta S}$, is much larger than the mean free path $l_{mfp}$. For example, for ambient air, $l_{mfp}$ is ~66 nm [19] and the intermolecular distance is ~3.4 nm (calculated from ideal gas pressure); if we look at the average pressure on a surface element $\Delta S$ in the order of μm$^2$, *i.e.*, $l_s$~μm, the above condition is satisfied.

First, we examine the pressure on a flat plate at rest in ambient air. Taking a *macroscopically* small surface element $\Delta S = \Delta x' \Delta y'$ in the plate (with $l_s \gg l_{mfp}$), for molecules with a given $V_z' > 0$ we construct the zeroth VE as the space region within a distance $h = V_z' \Delta t$ away from $\Delta S$ (Fig. 7a), which is filled with air molecules with $V_z'$ fixed but $V_x'$ and $V_y'$ unrestricted. Since pressure is a physical quantity averaged over certain time interval as well as the surface area $\Delta S$, we can choose a suitable $\Delta t$ so that $h = V_z' \Delta t \gg l_{mfp}$; furthermore, we divide the zeroth VE into N grid cells (Fig. 7a) and each grid cell has a width $W = V_z' \delta t \leq l_{mfp}$ so that intermolecular scattering within a grid cell is small. For a flat plate at rest, the statistical velocity distribution in the Nth grid cell (furthest away from the plate) is the Boltzmann distribution filled up to a cutoff speed $V_c$ (Fig. 3a); also, the distribution is the same in the 1st grid cell (in contact with the plate) since elastic scattering simply reverses the semi-sphere of incident velocity distribution with $V_z' > 0$ into the semi-sphere of outgoing velocity distribution with $V_z' < 0$ (Fig. 3a). Therefore, the statistical velocity distribution for each grid cell is the same, *i.e.*, the velocity distribution is statistically homogenous if performing translation in position space. With that, by taking snap shot in a time interval of $\delta t$ for the statistical distribution of all grid cells within the zeroth VE, we find that it can be treated *effectively* as if: **(1)** all the molecules in the 1st grid cell were able to reach the surface element $\Delta S$ within a time interval $\delta t$ (following the same reasoning in Sec. **II.2**) , and **(2)** all the molecules in the *n*th grid cell were able to move left to its nearest-neighbor (*n*-1)th grid cell closer to the plate within $\delta t$ (considering the statistical homogeneity discussed above). Therefore, after taking N such snap shot to finish a time interval of $\Delta t = N\delta t$, we reach the result that **it is *effectively* as if all molecules in the zeroth VE associated with a given $V_z' > 0$, defined as the space region within a distance $h = V_z' \Delta t$ away from $\Delta S$, were able to reach the surface element $\Delta S$ within a time interval $\Delta t$**. Consequently, the pressure on a plate at rest in ambient air with intermolecular scattering is obtained by putting $V_0 = 0$ in eq. (7)-(9), and thus we have $p_{net} = 0$ and $p_+ = p_- = p_0$ , with $p_0$ the static pressure for ideal gas. This explains why the experimentally observed static pressure for ambient air is so close to the pressure for ideal gas $p_0 = nk_BT$ (derived without considering intermolecular scattering), despite the fact that ambient air is by no means ideal but instead with significant intermolecular scattering (with a relatively short mean free path ~66 nm).

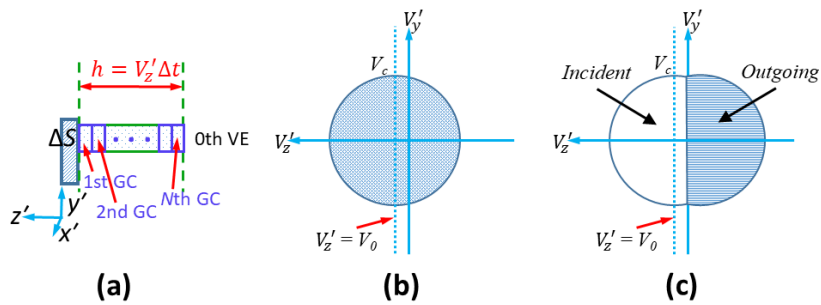

**Fig. 7.** (a) Schematic diagram of a surface element $\Delta S$ in a flat plate with the zeroth VE constructed for a given $V_z'$ and N grid cells (GCs) inside the zeroth VE. (b) The statistical velocity distribution for every grid cell at time $t = 0$, and for applicable grid cells at a later time as described in main text. (c) Instantaneous statistical velocity distribution in the 1st grid cell at $t = \delta t$.

Next, let the plate start moving at time $t = 0$ with a velocity $\boldsymbol{V_0}$ perpendicular to the plate (Fig. 2a). At time $t = 0$, the statistical velocity distribution for each grid cell in Fig. 7a is still the same (as described by Fig. 7b in the plate frame), *i.e.*, the velocity distribution is still statistically homogenous if performing translation in position space. At time $t = \delta t$, it can be treated *effectively* as if: **(1)** all the molecules in the 1st grid cell at $t = 0$ were able to reach the surface element $\Delta S$ (following the same reasoning in Sec. **II.2**) and elastically scattered by the plate; and **(2)** all the molecules in the *n*th grid cell at $t = 0$ were able to move left to its nearest-neighbor (*n*-1)th grid cell closer to the plate (considering the statistical homogeneity still held at $t = 0$). However, as a result of the elastic scattering, the instantaneous statistical velocity distribution in the 1st grid cell at $t = \delta t$, shown in Fig. 7c, is now different than that of other grid cells as shown in Fig. 7b. Subsequently, at $t = 2\delta t$, it is effectively as if: **(1)** another round of elastic scattering occurred at the plate for all the molecules with $V'_z > 0$ in the 1st grid cell at $t = \delta t$; **(2)** all the molecules with $V'_z < 0$ in the 1st grid cell at $t = \delta t$ (given by the instantaneous statistical velocity distribution of Fig. 7c) moved right to the 2nd grid cell within which the velocity distribution would be modified due to intermolecular scattering; and **(3)** the statistical velocity distribution for the *n*th grid cell with $n > 2$ is still the same as that of Fig. 7b. So on and so forth, as time passes in step of $\delta t$, similar physical processes would occur. Consequently, a steady state will be eventually reached so that the first few grid cells closer to the plate, *e.g.*, 5-10 grid cells considering the exponential dependence of relaxation in the form of $\exp(-d/l_{mfp})$ with $d$ the distance away from the plate surface, have velocity distribution between that of Fig. 3b and Fig. 3c. In the steady state, the velocity distribution in the 1st grid cell should be similar to that of Fig. 3c but with a new effective offset velocity $\boldsymbol{V_{e0}}$ with $0 < V_{e0} < V_0$ for the incident molecules, and a new effective temperature $T_{e0}$. With that, following Sec. **II.2**, the pressure on the plate can be expressed by eq. (4)-(6) but with $V_0$ and $T$ therein replaced by $V_{e0}$ and $T_{e0}$.

Similarly, if the plate is moving with an angle of attack in air with intermolecular scattering, eq. (7)-(9) can be applied to obtain the pressure by replacing $V_0$ and $T$ therein by $V_{e0}$ and $T_{e0}$, the effective offset speed and the effective temperature in the 1st grid cell closest to the plate surface. This intriguing phenomenon renders the result of this work to be of broad practical use for aerodynamic applications.

The above analysis can be tested experimentally by measuring lift/drag force of flat plate as a function of angle of attack at fixed speed in ambient air and fitting the data with eq. (9)-(11) to extract the effective $V_{e0}$ and $T_{e0}$.

Theoretically, this effect can be investigated quantitatively in future by applying the direct simulation Monte Carlo method [11-12] to simulate intermolecular collisions inside the zeroth VE of Fig. 7a, using a cyclic boundary condition between neighboring VEs, *i.e.*, molecules going out to 1st VE is treated as coming back from the other side of -1st VE (considering the symmetry of velocity distribution as discussed in Sec. **II.3**.), and *vice versa*. Alternatively, one can also theoretically estimate $V_{e0}$ and $T_{e0}$ by treating the rebounding of molecules at the plate surface as a perturbation and using the linear response theory for the *diffusive regime*.[17]

## III. Friction on moving airfoils with surface roughness

## III.1. Friction on a plate with surface roughness moving along a direction parallel to the plate plane

In Sec. **II**, we deal with ideally flat plates without surface roughness, e.g., atomically flat crystal surfaces or molecular-level flat surfaces formed by adsorption of molecules, for which pressure or normal force is induced as a result of elastic scattering of molecules at the air-plate interface. Next, we show that for flat plates with surface roughness, frictional force parallel to the plate plane can also be obtained via the Volume-Element method considering elastic scattering of molecules.

First, we consider a flat plate with a single step of protrusion from the plate surface $S$, moving with a constant velocity $V_0$ parallel to the plate plane in the laboratory frame (Fig. 8a). The surface protrusion forms a perpendicular side wall out of the plate characterized by a side-wall surface area $\Delta S$, and we assume that such a side wall is ideally flat in atomic or molecular level, e.g., a crystalline protrusion.

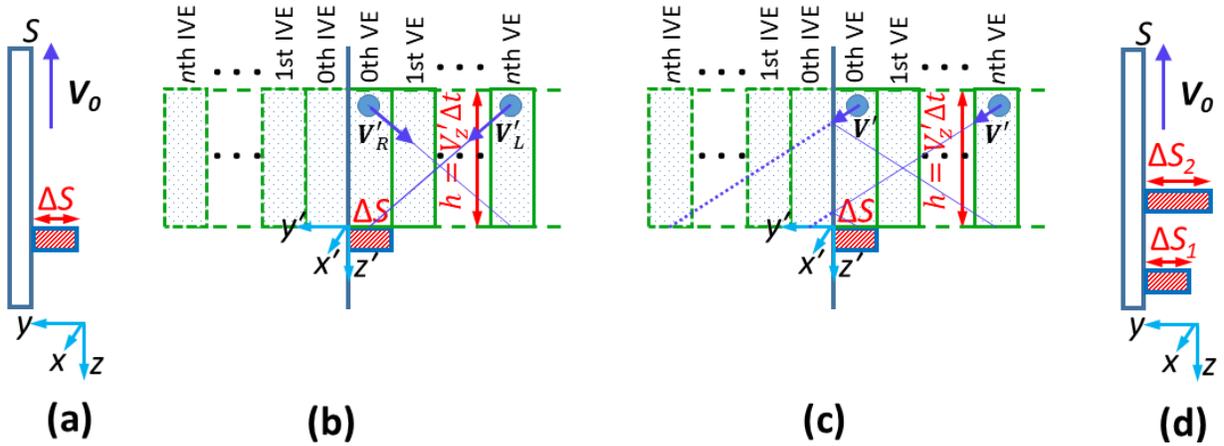

**Fig. 8.** Schematic diagrams of a flat plate $S$ moving with a velocity $V_0$ parallel to the plate plane: **(a)** in the laboratory frame with a single step of protrusion $\Delta S$; **(b-c)** in a reference frame fixed at the plate with a set of VEs and IVEs constructed for a given $V_z'$ and used for analysis of molecular motion; and **(d)** in the laboratory frame with multiple steps of protrusion $\Delta S_i$.

Again, we work in the reference frame fixed at the moving plate. For molecules with a given $V_z' > 0$ we construct the zeroth VE as the space region within a distance $h = V_z'\Delta t$ away from $\Delta S$ (Fig. 8b); by periodic translation of the surface element $\Delta S$ and its associated VE to the right of the plate plane, a set of VEs indexed by positive integers are formed (Fig. 8b). Furthermore, we define image volume elements (IVEs) as the mirror images of corresponding VEs with respect to the plate plane (Fig. 8b), *i.e.*, nth IVE is the mirror image of nth VE. The introduction of IVEs makes it convenient to analyze the effect of those molecules scattered by the plate plane.

Owing to the symmetry of the probability distribution and the mirror reflection of molecules at the plate plane (Figs. 7b and 7c), the total momentum transfer on the protruded side-wall surface element $\Delta S$, induced by those molecules physically incident from all VEs, is equal to that induced by all molecules contained in the zeroth VE (no matter if the molecules are actually travelling out of the zeroth VE or not). This argument is supported by the fact that all the molecules moving inside the zeroth VE can be classified into three cases:

**(1) Molecules right-going away from the zeroth VE and reaching an extended side-wall surface element out of the zeroth VE**. Due to the symmetry of the probability distribution

shown in eq. (*2*) with $V_0$ antiparallel to the $z'$ axis, for any molecule with velocity $V'_R$ right-going away from the zeroth VE to the *n*th VE there must exist a *statistically* pairing molecule with velocity $V'_L$ left-going from the *n*th VE to the zeroth VE (Fig. 8b) where $V'_R$ and $V'_L$ have same $V'_z$ but inverted $V'_x$ and $V'_y$; and *vice versa* so that one-to-one correspondence is formed between the pairing right-going and left-going molecules.

**(2) Molecules left-going away from the zeroth VE but reaching an extended side-wall surface element out of the zeroth VE after being reflected by the plate plane.** For a molecule with velocity $V'$ left-going from the zeroth VE initially but being bounced back to reach the *n*th VE finally (Fig. 8c), it is as if the molecule were "effectively" reaching the *n*th IVE (indicated by extended dash line in Fig. 8c); since the velocity probability distribution of eq. (*2*) is translationally invariant in position space, for the above molecule with velocity $V'$ left-going from the zeroth VE but "effectively" reaching the *n*th IVE, there must exist a *statistically* pairing molecule with the same velocity $V'$ coming from the *n*th VE but "effectively" reaching the zeroth IVE (*i.e.*, actually reaching the zeroth VE due to reflection as shown in Fig. 8c); and *vice versa* so that one-to-one correspondence is again formed between the pairing molecules.

**(3) Molecules starting from the zeroth VE and reaching the side-wall surface element inside the zeroth VE, either with or without being reflected by the plate plane.**

Upon reaching the relevant side-wall surface element, each pairing molecule contributes the same amount of momentum transfer of $\Delta P = 2mV'_z$.

Therefore, as described in Sec. **II.2**, eqs. (4)-(6) are still applicable here to calculate the pressure acting on the protruded side-wall surface element $\Delta S$, and thus the frictional force $F_f$ parallel to the plate plane is obtained as:

$$F_f = p_{net} \cdot \Delta S, \qquad (15)$$

where $p_{net}$ is given by eq. (6).

Next, we consider a flat plate with a series of step protrusions sparsely distributed on the plate surface (Fig. 8d). Each surface protrusion forms a perpendicular side wall out of the plate characterized by a side-wall surface area $\Delta S_i$. In the case that the spacing between adjacent step protrusions is much larger than the mean free path $l_{mfp}$ characterizing intermolecular collisions, eq. (13) is still valid for calculating the frictional force on each side-wall surface element $\Delta S_i$ independently, and thus the total frictional force $F_f$ can be obtained as:

$$F_f = \sum p_{net} \cdot \Delta S_i, \qquad (16)$$

where $p_{net}$ is given by eq. (6).

**III.2. Friction on a plate with surface roughness moving with angle of attack**

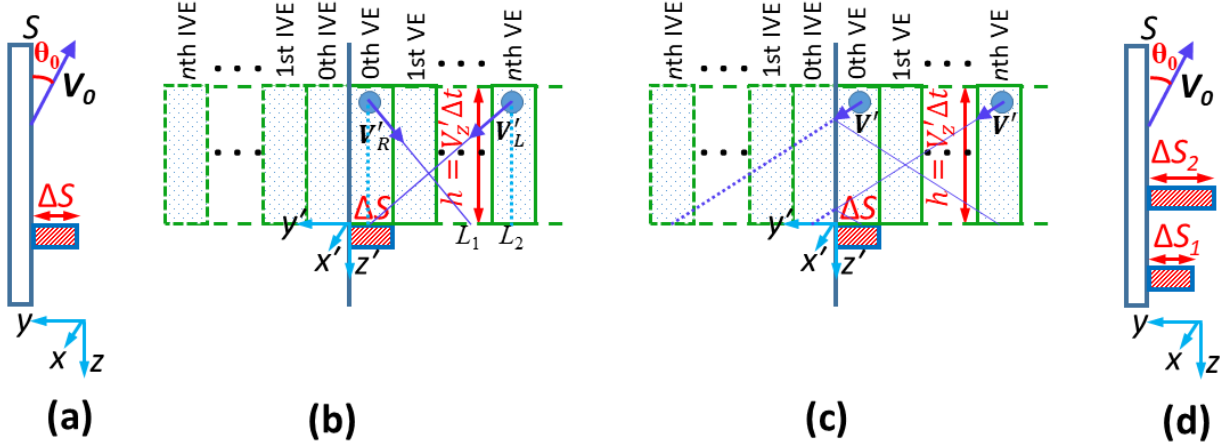

**Fig. 9.** Schematic diagrams of a flat plate $S$ moving with a velocity $V_0$ at an angle $\theta_0$ from the plate plane: **(a)** in the laboratory frame with a single step of protrusion $\Delta S$; **(b-c)** in a reference frame fixed at the plate with a set of VEs and IVEs constructed for a given $V_z'$ and used for analysis of molecular motion; and **(d)** in the laboratory frame with multiple steps of protrusion $\Delta S_i$.

We again start by a flat plate with a single step of protrusion from the plate surface $S$, moving in the laboratory frame with a constant velocity $V_0$ at an angle $\theta_0$ from the plate plane (Fig. 9a); the perpendicular step protrusion is characterized by a side-wall surface area $\Delta S$. As described in Sec. **III.1**, working in the reference frame fixed at the moving plate, for molecules with a given $V_z' > 0$ we construct a set of VEs and IVEs within a distance $h = V_z'\Delta t$ away from $\Delta S$ (Figs. 8b and 8c).

Similarly, all molecules moving inside the zeroth VE for a given $V_z'$ can be classified into those three cases outlined in Sec. **III.1**. For Case (2) and Case (3), we can directly follow the analysis in Sec. **III.1**. However, the previous analysis of Sec. **III.1** is not directly applicable to the situation of Case (1) here since the one-to-one correspondence between the pairing molecules incoming and outgoing from the zeroth VE is broken, as detailed below.

**Case (1): Molecules right-going away from the zeroth VE and reaching an extended side-wall surface element out of the zeroth VE**. We consider any given molecule initially located in the zeroth VE (Fig. 9b) with arbitrary coordinates $(x_R', y_R', z_R')$ but right-going with velocity $\boldsymbol{V_R'} = V_{Rx}'\boldsymbol{i} + V_{Ry}'\boldsymbol{j} + V_Z'\boldsymbol{k}$ and it reaches the extended side-wall surface element out of the zeroth VE after travelling a distance of $L_1 = |V_{Ry}'| \cdot \delta t$ in $y'$ direction with $\delta t = |z_R'| / V_Z'$. Here $V_{Ry}' = V_{Ry} + V_0 \sin\theta_0 < 0$ is the velocity component in the plate frame, with $V_{Ry}$ being the component in the laboratory frame.

For any of the above right-going molecule, there exists a *statistically* pairing molecule, located in the $n$th VE (Fig. 9b) with coordinates $(x_R' + D, y_R' - L_2, z_R')$, but left-going with velocity $\boldsymbol{V_L'} = -V_{Rx}'\boldsymbol{i} + V_{Ly}'\boldsymbol{j} + V_Z'\boldsymbol{k}$, which is able to reach the surface element of the zeroth VE at the position $(x_R', y_R', 0)$ by satisfying: $V_{Ly}' = -V_{Ry} + V_0 \sin\theta_0 > 0$, $L_2 = V_{Ly}' \cdot \delta t$, and $D = V_{Rx}' \cdot \delta t$. In the laboratory frame, the above right-going and left-going molecules have the same $V_z$ component, but opposite $V_x$ and $V_y$ components; thus their distribution probability is the same according to the symmetry of the Boltzmann distribution in eq. (*1*). We note that $L_1 < L_2$ always holds, since for right-going molecule

$V_{Ry}' = V_{Ry} + V_0 \sin\theta_0 < 0$,

so that $V_{Ry} < -V_0 \sin\theta_0 < 0$, and then we have $|V_{Ry}'| < V_{Ly}'$ using

$$\left|V_{Ry} + V_0 \sin\theta_0\right| < |V_{Ry}| + |V_0 \sin\theta_0|.$$

On the other hand, for any left-going molecule started from the *n*th VE but travelling a distance of $L_2$ in $y'$ direction to reach the surface element of the zeroth VE, there also exists a *statistically* pairing molecule started from the zeroth VE but travelling a distance of $L_1$ in $y'$ direction to reach an surface element *either inside or outside* the zeroth VE (the distance ratio is: $L_1/L_2 = |V'_{Ry}|/V'_{Ly}$, always less than 1 from previous analysis). However, in certain cases when $L_2$ extends out of the zeroth VE but $L_1$ is still within the zeroth VE (*i.e.*, $L_1$ is less than the length scale characterizing $\Delta S$), the number of molecules flowing into the zeroth VE and scattered by the surface element $\Delta S$ is larger than that flowing out of the zeroth VE; this causes an imbalance of incoming and outgoing molecules for the zeroth VE and breaks the one-to-one correspondence between the pairing molecules.

Nevertheless, since the surface element $\Delta S$ characterizing surface roughness is typically small, the effect of such slight imbalance can be neglected, and as a good approximation, the total momentum transfer on the protruded side-wall surface element $\Delta S$, induced by those molecules physically incident from all VEs, is equal to that induced by all molecules contained in the zeroth VE (no matter if the molecules are actually travelling out of the zeroth VE or not).

Therefore, as described in Sec. **II.3**, eqs. (7)-(9) are still applicable here to calculate the pressure acting on the protruded side-wall surface element $\Delta S$, and thus the frictional force $F_f$ parallel to the plate plane is obtained as:

$$F_f = p_{net} \cdot \Delta S, \tag{17}$$

where $p_{net}$ is now given by eq. (9).

Next, we consider a flat plate with a series of step protrusions sparsely distributed on the plate surface (Fig. 9d), each forming a perpendicular side wall out of the plate characterized by $\Delta S_i$. Again, if the spacing between adjacent step protrusions is much larger than the mean free path $l_{mfp}$ characterizing intermolecular collisions, eq. (15) can be used for calculating the frictional force on each $\Delta S_i$ independently, and the total frictional force $F_f$ is:

$$F_f = \sum p_{net} \cdot \Delta S_i, \tag{18}$$

where $p_{net}$ is given by eq. (9).

As shown above, frictional force is found to be a direct consequence of elastic scattering of air molecules due to the surface roughness of a plate airfoil. In addition, inelastic scattering of molecules by a plate (even if it is ideally flat) may contribute to part of the frictional force, *e.g.*, via phonon excitation in solid or surface adsorption/desorption of air molecules. However, the probability for inelastic scattering should be small compared with elastic scattering and it is reasonable to assume that surface roughness is the dominant source for friction. When a plate with surface roughness is moving with angle of attack (Fig. 9), the frictional force is parallel to the plate surface and contributes to both lift and drag.

Below we further discuss the relation of friction in this study to the theory of viscosity,[1, 20] in terms of the origin of the "no-slip" boundary condition and the boundary layer.

First, a basic assumption for viscous flow is the "no-slip" boundary condition: *i.e.*, the tangential component of the relative velocity between a viscous fluid and a solid object is assumed to be zero at the fluid-solid interface. By examining the center-of-mass motion for the statistical assembly of molecules scattered at the air-solid interface for a plate with surface roughness, we can investigate the validity and the limit of the no-slip boundary condition as a dependence on the characteristics structure of surface roughness. For example, for a plate with surface roughness in the form of side-wall surface protrusions perpendicular to the plate plane (Fig. 8d), if the plate starts moving at $\boldsymbol{V}_0$ parallel to the plate, the velocity distribution of molecules at the plate surface is disturbed instantaneously due to elastic scattering between molecules and the plate: as described in Figs. 3b and 3c, the portion of incident velocity distribution (unshaded area) is mirror reflected into the portion of outgoing velocity distribution (shaded area) due to elastic scattering, and the incident and outgoing portions together form the complete instantaneous velocity distribution for molecules at the plate surface. If the fluid is in *diffusive regime* (*i.e.*, viscous), intermolecular collisions cause the evolution of the instantaneous distribution of Fig. 3b or 3c into a thermalized quasistatic-equilibrium state described by Fig. 3f, resulting in an effective center-of-mass displacement velocity $\boldsymbol{V}_e = \boldsymbol{V}_0$ for the fluid at the plate surface; thus the "no-slip" boundary condition occurs. However, if for a real plate where the orientation angle between the side-wall surface protrusions and the plate plane is randomly distributed, the effective displacement velocity $\boldsymbol{V}_e$ for the fluid at the plate surface should be a statistical average of all surface protrusions and then it is more reasonable to take the boundary condition as $\boldsymbol{V}_e = \xi \boldsymbol{V}_0$ with an attenuation factor $0 < \xi \leq 1$.

Second, the origin of a thin boundary layer important for Prandtl's viscosity theory [1, 20] also can be explained by relaxation of the quasistatic-equilibrium state with an effective $\boldsymbol{V}_e = \xi \boldsymbol{V}_0$ for the fluid at the plate surface resultant from the surface roughness as discussed above. In brief, due to intermolecular scattering with mean free path $l_{mfp}$, the fluid flow at a distance $r$ away from the plate surface should have a spatially damping profile proportional to $\exp(-r/l_{mfp})$. This exponential spatial dependence is short-ranged and thus naturally gives a thin boundary layer around the moving plate, within which the fluid-flow speed is rapidly damping.

### IV. Summary and Future Directions

In summary, based on Statistical Mechanics and elastic scattering at the air-solid interface, we have implemented the Volume-Element method to address the aerodynamic lift and drag, and obtained analytical expressions for the pressure on canonical flat plates (Sec. **II**) and friction on flat plates with surface roughness (Sec. **III**). If the effective temperature around a plate has a spatial dependence, Monte Carlo simulations can be included to numerically evaluate the pressure on the plate (Sec. **II.2**). In general, the Volume-Element approach can be applied to numerically evaluate pressure-induced lift and drag on any convex-shape airfoil (Sec. **II.4**).

For concave-shape airfoils, the situation is more complicated and requires further investigations. For an airfoil with arbitrary concave shape at rest in the laboratory frame, the Volume-Element method can still be directly applied by constructing VEs exactly following the surface profile of the airfoil; due to the rotational and inversion symmetry in velocity space and translational invariance in position space for the probability distribution, the pressure is found to be the well-known static pressure for ideal gas. If a concave air foil is moving at velocity $\boldsymbol{V_0}$, symmetry in the probability distribution is broken in velocity space in the reference frame fixed at the airfoil, and thus the Volume-Element method is not directly applicable any more in general cases except for some specific concave shape where certain symmetry in probability distribution is conserved. However, if combined with Monte Carlo simulations to capture molecular motions in enough number of VEs, it may be also viable to numerically obtain the pressure for an arbitrary-shape concave airfoil; particularly, for air with intermolecular scattering and $l_s \gg l_{mfp}$, only the zeroth VE need to be considered (Sec. **II.5**).

In future works, interactions between molecules can also be included to address lift and drags in denser fluids. In addition, similar method can be developed to address pressure induced by Boson or Fermi gases: *e.g.*, pressure on moving objects in optical fields (photon gases), and electromigration effect in narrow electrical conductors (electron gases).